\documentclass[a4paper,aps,amsmath,amssymb,showpacs,10pt,twocolumn]{revtex4}

\usepackage[dvips]{graphicx}
\usepackage{braket}
\usepackage{bbm}
\usepackage{amssymb}
\usepackage{amsmath}
\usepackage{amsthm}
\DeclareMathOperator{\Prob}{Prob}

\newtheorem{theorem}{Theorem}[section]
\newtheorem{proposition}[theorem]{Proposition}

\newtheorem{lemma}[theorem]{Lemma}

\begin{document}

%==========================================================================================
% TITLE PAGE
%==========================================================================================

\title{Fluctuation symmetries for work and heat}
\author{Marco Baiesi}
\author{Tim Jacobs}
 \email[Corresponding author:~]{tim.jacobs@fys.kuleuven.be}
\author{Christian Maes}
\author{Nikos S. Skantzos}
 \affiliation{Instituut voor Theoretische Fysica, Celestijnenlaan 200D, K.U.Leuven, B-3001 Leuven, Belgium}

\begin{abstract}\noindent
We consider a particle dragged through a medium at constant
temperature as described by a Langevin equation with a
time-dependent potential. The time-dependence is specified by an
external protocol.  We give conditions on potential and protocol
under which the dissipative work satisfies an exact symmetry in
its fluctuations for all times.  We also present counter examples
to that exact fluctuation symmetry when our conditions are not
satisfied. Finally, we consider the dissipated heat which differs
from the work by a temporal boundary term.  We explain
why there is a correction to the standard fluctuation theorem
due to the unboundedness of that temporal boundary. However, the corrected
fluctuation symmetry has again a general validity.
\end{abstract}

\pacs{05.10.Gg, 05.40.-a, 05.70.Ln}
\maketitle

% ===================================================================
% INTRODUCTION
% ===================================================================

\section{Introduction}
Recent years have seen an explosion of results and discussions on
a particular symmetry in the fluctuations of various dissipation
functions.  While started in the context of smooth dynamical
systems and of thermostating algorithms and
simulations \cite{ev, GC}, soon the symmetry was judged relevant in
the construction of nonequilibrium statistical mechanics.
Moreover, a unifying scheme was developed under which the various
fluctuation symmetries were found to be the result of a common
feature. The basic idea is there that a dissipation function for a
physical model can be identified with the source of time-symmetry
breaking in the statistical distribution of system histories, see
e.g. \cite{poincare, mn} for more details.    That dissipation
function is mostly related to the variable entropy production but,
depending on the particular realization, can also refer to heat
dissipation or to dissipative work. For a given effective model,
one of course always needs to check again that basic relation
between time-reversal and
 dissipation.\\

  In the
present paper, we look at a particle's position $x_t$ that
undergoes a Langevin evolution for a time-dependent potential
$U_t$. Because of that time-dependence, which is externally
monitored, work $W$ is done on the particle. At the same time,
some of it flows as heat $Q$ to the surrounding medium, checking
the conservation of energy $\Delta U = U_\tau(x_\tau) - U_0(x_0) =
W - Q$ for the evolution during a time interval $0\leqslant t
\leqslant \tau$. Both $W$ and $Q$ are fluctuating quantities and
they are path-dependent. Our main result concerns a symmetry in
the fluctuations of $W$. We give conditions on the potential and
on its time-dependence $U_t$ under which the fluctuation symmetry
for $W$ is exact, i.e., that for all times $\tau$, without further
approximation,
\begin{equation}
\label{EFT-work} \frac{\Prob_{\rho_0}\left( W^{dis} =
w\right)}{\Prob_{\rho_0}\left( W^{dis} = -w\right)}
 = \exp\left(\beta w\right)
\end{equation}

Here,  the particle's position is initially distributed with $\rho_0
\sim \exp(-\beta U_0)$ according to thermal equilibrium at inverse
temperature $\beta$. The notation $W^{dis}$ refers to the dissipated
work \eqref{eq:definition-diswork} which equals the work $W$ up to
a difference in free energies. If the evolution would be
reversible, then $W^{dis}=0$. In general, and confirming the second
law, we have $\langle W^{dis} \rangle \geqslant 0$ but \eqref{EFT-work} also
takes into account the trajectories where $W^{dis}<0$. The exact
fluctuation symmetry \eqref{EFT-work} tells us that such ``negative
dissipative work'' trajectories are exponentially damped.\\

Since the heat $Q$ differs from the dissipated work only by a
temporal boundary term $\Delta$, $Q = W^{dis} - \Delta$ where
$\Delta=\Delta(\tau;x_0,x_\tau)$ is non-extensive in time $\tau$,
one could perhaps expect that $Q$ satisfies the standard
fluctuation symmetry, i.e., that the same as in \eqref{EFT-work}
is true after taking the logarithm and letting $\tau\uparrow
+\infty$:
\begin{equation}
\label{FS-heat}
\lim_{\tau\uparrow +\infty} \frac 1{\tau} \log
\frac{\Prob_{\rho_0}\left( Q = q \tau\right)}{\Prob_{\rho_0}\left(
Q = -q \tau \right)} \stackrel{?}{=} \beta q
\end{equation}
for $q$ the heat per unit time.
Interestingly, that is not what always happens, see
\cite{CohenVanZon}. We will explain how the unboundedness of the
potential $U_\tau$ can change the symmetry \eqref{FS-heat}. For
small enough $q$ (basically, for $0\leq q\tau \leq \langle W^{dis}
\rangle$) the relation \eqref{FS-heat} remains intact but for all
large enough $q$ the lefthand-side of \eqref{FS-heat} saturates
and is constant.\\

In what follows, we discuss the symmetry relations \eqref{EFT-work} and
\eqref{FS-heat} in mathematical detail. In particular, we give
 near to optimal conditions on
potential and protocol for which \eqref{EFT-work} holds.
  Before, that was shown only for the case of a harmonic potential where the minimum
of the potential is moved with a fixed speed via an explicit
calculation  \cite{CohenVanZon}.  There it was also found that
\eqref{FS-heat} can be broken and the modification was explicitly
calculated.  Here we will argue for more general protocols and
potentials to give estimates about the range of validity of
\eqref{FS-heat}.  The main point is to understand when and how
terms, non-extensive in the time $\tau$, can still contribute to
the large deviations of the heat $Q$.

\section{Model and results}

In the present paper we apply the general scheme and algorithm of
\cite{poincare,mn} to a model that has previously and recently been
considered by a number of groups
\cite{stuttgart,evans,CohenVanZon,wang}. We find optimal conditions
on potential and protocol under which the dissipated work satisfies
an exact fluctuation symmetry, i.e., one that is valid for all
times.  The heat differs from that dissipated work via a temporal
boundary term and also satisfies some general fluctuation symmetry
asymptotically in time.  Because the potential is unbounded, that last
symmetry is not the same as in the standard steady state fluctuation
theorem.  Below we give more details.

\subsection{Model}

We consider a family of one-dimensional potentials $U_t, t\in
[0,\tau]$, as parameterized via a deterministic protocol $\gamma_t$:
$U_t(x) = U(x,\gamma_t)$, with $x, \gamma_t \in \mathbbm{R}$. The corresponding equilibria at inverse
temperature $\beta$ are
\begin{eqnarray}
\label{eq:definition rho}
\rho_t(x) & \equiv & \frac{e^{- \beta U_t(x)}}{Z_t} \\
Z_t & \equiv & \exp[-\beta F_t] \equiv
\int_{-\infty}^{+\infty} dx \,\, e^{- \beta U_t(x)} \nonumber
\end{eqnarray}
with Helmholtz free energy $F_t$. The time-dependence in
$\gamma_t$ is  supposed to be given and can be quite arbitrary; of
course the partition function $Z_t$ must be finite. The dynamics
is now specified by a Langevin-It\^o--type equation
\begin{equation}\label{lang}
d x_t = -\frac{\partial U_{t}}{\partial x} (x_t) \,\, dt +
\sqrt{\frac{2}{\beta}} \,\, d b_t
\end{equation}
where $db_t$ is standard white noise. Such dynamics have been
considered before in a wide variety of contexts but for
fluctuation theorems the emphasis has been on the Gaussian case.
An experimental realization \cite{wang} of that dynamics was
theoretically investigated by \cite{CohenVanZon}, who started from
\eqref{lang} with
\begin{equation}
\label{eq:potential-CohenVanZon} U_t(x) = \frac{(x - vt)^2}{2}
\end{equation}
A more general analysis for driven harmonic diffusive systems
was given in \cite{SpeckSeifert}. Quite recently in \cite{stuttgart} further experiments
were considered for more general potentials and protocols.\\

In the present paper we work with the general \eqref{eq:definition
rho} but we sometimes restrict ourselves to the physically
relevant case of
\begin{equation}\label{gen}
U_t(x) = U(x - \gamma_t)
\end{equation}
for a given protocol $\gamma = (\gamma_t, t\in [0,\tau])$ that
marks the shift in a potential $U$ as time goes on.  In that case
$F_t$ does not depend on time and the associated difference in
free energies
\begin{equation}
\label{eq:deltaF} \Delta F = -\frac
1{\beta}\ln\left(\frac{Z_{\tau}}{Z_0}\right)
\end{equation}
is zero.\\

 The model \eqref{lang} defines a Markov
diffusion process.  We write
\begin{displaymath}
\omega = (x_t, t\in [0,\tau])
\end{displaymath}
for the (random) positions of the particle.  If the initial
distribution of the position $x_0$ is given via a density $\rho$,
then $\Prob_\rho(\omega|\gamma)$ denotes the probability density
of observing a trajectory $\omega$ under the influence of the
protocol $\gamma$, with respect to the thermal noise
$\sqrt{2/\beta}\, db_t$. Given a path $\omega$ and a protocol
$\gamma$ we also consider their time-reversed versions:
\begin{equation}
\label{eq:def-timereversal}
\begin{aligned}
\Theta \omega_i &\equiv \omega_{\tau - i} = x_{\tau - i}\\
\Theta \gamma_i &\equiv \gamma_{\tau - i}
\end{aligned}
\end{equation}
%We will write $\Theta U_t$ for the potential under the
%time-reversed protocol $\Theta \gamma$.

\subsection{Problem}

The observables of interest are the work and the heat.\\

  The work
$W_\gamma$ is associated to the external agent, in changing the
potential via the protocol $\gamma$,
\begin{equation}
\label{eq:definition-work} W_{\gamma}(\omega) \equiv \int_0^{\tau}
dt \,\, \dot{\gamma}_t \, \frac{\partial U_t}{\partial
\gamma_t}(x_t)
\end{equation}
The dissipative work can then be identified with
\begin{equation}
\label{eq:definition-diswork} W_{\gamma}^{dis}(\omega) =
W_{\gamma}(\omega) - \Delta F
\end{equation}
One has to remember here that for a reversible and isothermal
evolution the change in free energy precisely equals the work
$W_\gamma$ done on the system.  Furthermore, in the situation \eqref{gen}
one has $\Delta F = 0$ so that $W_\gamma = W_{\gamma}^{dis}$.\\

The heat $Q_\gamma$ is most easily defined via the first law
of thermodynamics.
\begin{displaymath}\label{1law}
\Delta U = U_{\tau}(x_{\tau}) - U_{0}(x_{0}) = W_{\gamma}(\omega)
- Q_{\gamma}(\omega)
\end{displaymath}
\begin{equation}
\label{eq:definition-heat} Q_{\gamma}(\omega) \equiv -
\int_0^{\tau} dx_t
 \circ \frac{\partial U_t}{\partial x}(x_t) = - \int_0^{\tau} dt \,\,
\dot{x}_t \, \frac{\partial U_t}{\partial x}(x_t)
\end{equation}
The integral (with the ``$\circ$'') should be understood in the
sense of Stratonovich; it coincides better with the usual
intuition of integrals and it does not suffer from the lack of
time-symmetry in the It\^o integral which will be important for
us, see also
\cite{Mortensen}.\\

In the present paper we ask
\begin{enumerate}
\item under what conditions the work
\eqref{eq:definition-work} or \eqref{eq:definition-diswork}
satisfies an exact fluctuation symmetry (EFS) \eqref{EFT-work}.
\item what are the possible corrections to the
standard fluctuation symmetry \eqref{FS-heat} for the fluctuations of
the heat \eqref{eq:definition-heat}.
\end{enumerate}

So far, these questions have been theoretically investigated via
explicit computation for the special case of a linearly dragged
particle in the harmonic potential
\eqref{eq:potential-CohenVanZon}, in \cite{CohenVanZon}; question
(2) has been generally addressed in \cite{galla}. Experimental and
numerical work (in agreement with the results below) was done in
\cite{stuttgart,stuttgart2,evans,wang}.

\subsection{Results}
\label{ssec:results}

We start with the exact fluctuation symmetry for the work
\eqref{eq:definition-work}. First we consider the harmonic case
$U(x) = x^2/2$ with a general protocol $\gamma_t$ as in \eqref{gen}.
Then we give a general condition under which the work satisfies an
EFS, and we give instances under which that condition is
satisfied. Counter examples (for which the work does not satisfy an
EFS) will show why these conditions are close to optimal.  We end
with a discussion on the relevance of temporal boundary terms in
the large deviations of the heat \eqref{eq:definition-heat}. For the proofs
we refer to section \ref{ssec:proofs}.

\subsubsection{Work}

First look at quadratic potentials, e.g. for
\begin{equation}
\label{eq:potential-quadratic}
U_t(x) = \frac{(x - \gamma_t)^2}{2}
\end{equation}
which coincides with the potential
\eqref{eq:potential-CohenVanZon} if $\gamma_t = vt$. For that
class of quadratic potentials, as in
\eqref{eq:potential-quadratic}, one has a Gaussian distribution of
the work \eqref{eq:definition-work} for all protocols
$\gamma_t$.\\

In what follows the probability density for the (dissipated) work
is denoted by $\Prob_{\rho_0}\big(W_{\gamma}^{(dis)}(\omega) =
w\big)$ as a function of $w\in \mathbbm{R}$. This (dissipated) work $W_{\gamma}^{(dis)}$
depends on the time $\tau$, see \eqref{eq:definition-work} - \eqref{eq:definition-diswork}.
\begin{proposition}[Harmonic case]~\\
\label{prop:harmonic}
If the distribution of the work $W_{\gamma}$ is Gaussian,
then for a general protocol $\gamma_t$ in \eqref{gen}:
\begin{align}\label{har}
\frac{\Prob_{\rho_0}\big(W^{dis}_{\gamma}(\omega) =
w\big)}{\Prob_{\rho_0}\big(W^{dis}_{\gamma}(\omega) = - w\big)}
 & = \exp(\beta w)
\end{align}
for all times $\tau$.
\end{proposition}
That fluctuation symmetry is easily checked to hold also for
quadratic potentials which are more general than
\eqref{eq:potential-quadratic}. One could argue that any
Gaussian distributed observable can be made to satisfy a
fluctuation theorem by rescaling the mean and the variance.
However, that is not what happens here: no scaling at all is required
for the work $W_{\gamma}$ to satisfy the exact fluctuation symmetry.\\

For more general potentials, we start by specifying a general condition:\\

\textbf{Assumption:}
We assume that there exists an
involution $s$ on path-space,  $s^2 = \mathbbm{1}$, with $s\Theta
= \Theta s$ and such that
\begin{equation}
\label{eq:definition_s}
\Prob_{\rho_{\tau}} ( \omega | \Theta \gamma ) = \Prob_{\rho_0} ( s \omega | \gamma )
\end{equation}

The involution $s$ relates trajectories under the protocol
$\gamma$ and its time-reversed protocol $\Theta \gamma$. The next
theorem stipulates that the existence of $s$ implies an exact
fluctuation theorem for the work.  We illustrate that assumption
below by enumerating the cases where the assumption is certainly
verified, see also in Section \ref{disc}.

\begin{theorem}[EFS Work]~\\
\label{theorem:efs-work}
Under the assumption  \eqref{eq:definition_s} above, the
dissipative work \eqref{eq:definition-diswork} satisfies an EFS:
for all $\tau>0$ and for all $w$,
\begin{equation}
\label{eq:theorem_workEFT}
\frac{\Prob_{\rho_0}\left( W_{\gamma}^{dis}(\omega) =
w\right)}{\Prob_{\rho_0}\left( W_{\gamma}^{dis}(\omega) = -w\right)}
 = \exp\left(\beta w\right)
\end{equation}
\end{theorem}

The assumption \eqref{eq:definition_s} can be split in some two
subassumptions as we now state.
\begin{proposition}
\label{prop:cases}~\\ Suppose either $(i)$ that the protocol is symmetric
under time-reversal $\gamma_t=\gamma_{\tau-t}\equiv\Theta \gamma_t$, or $(ii)$ that the protocol is
antisymmetric $\gamma_t -
\gamma_0 = \gamma_\tau - \gamma_{\tau-t} \equiv - \Theta(\gamma_t -
\gamma_0) $ and that the potential $U$ obeys \eqref{gen} and is
symmetric, $U(x)=U(-x)$. Then assumption \eqref{eq:definition_s}
and hence the EFS \eqref{eq:theorem_workEFT} are verified.
\end{proposition}

The EFS for the harmonic case $U(x)=x^2/2$ with constant velocity
$\gamma_t=vt$ as in \eqref{eq:potential-CohenVanZon}, see
\cite{CohenVanZon,wang}, is treated by Proposition \ref{prop:harmonic} but is of
course also a special case of Proposition \ref{prop:cases}.\\

We will see further in Section \ref{ssec:numerics} how our
conditions are in fact optimal.  We can however already
observe here how some symmetry of the protocol must enter when dealing with
an arbitrary potential.  Consider indeed, if only formally, $U(x) =
x, x>0$ with a wall $U(x) = +\infty$ for $x\leq 0$ in
\eqref{eq:definition-work}.  We can then safely assume that the
trajectory satisfies $x_t -\gamma_t >0$ and
\eqref{eq:definition-work} gives that the work $W_\gamma =
W_\gamma^{dis} = \gamma_\tau - \gamma_0$.  Obviously this (constant)
expression never satisfies an EFS unless (and then trivially)
$\gamma_\tau=\gamma_0$.

% ===================================================================
% HEAT
% ===================================================================

\subsubsection{Heat}\label{heatresults}
The heat $Q_\gamma$ defined in \eqref{eq:definition-heat} equals
the dissipative work $W_\gamma^{dis}$ up to some temporal boundary
term:
\[
Q_\gamma = W_\gamma^{dis} + \Delta (F - U)
\]
The temporal boundary $\Delta (F - U)$ is, modulo the factor
$\beta$, the change of equilibrium entropy in going from the
equilibrium described by $\rho_0$ to that given by $\rho_\tau$.
For the fluctuations of the heat  we start from a situation where
we already have the EFS \eqref{eq:theorem_workEFT} for the
(dissipative) work.\\

We are here concerned with the situation where the potential in \eqref{gen} is unbounded and we
assume that for some $\varepsilon,v>0$
\begin{equation}\label{as2}
U(x) \geq |x|^{1+\varepsilon}, \gamma_t = vt \end{equation} at
least for $|x|$ and $t$ sufficiently large.
  For the average work we write
\begin{displaymath}
\lim_{\tau \rightarrow +\infty}\frac{\langle Q_\gamma
\rangle}{\tau} = \lim_{\tau \rightarrow +\infty} \frac{\langle
W_\gamma \rangle}{\tau} \equiv \overline{w}
\end{displaymath}
We further continue
to assume the well-defined dynamics \eqref{lang} with the EFS
\eqref{eq:theorem_workEFT} for the (dissipative) work.
  The latter can be summarized by introducing the rate function
  $I(w)$
  which, in logarithmic sense and asymptotically as $\tau\uparrow
  +\infty$, governs
  \[
\Prob\big(W_\gamma^{dis}=w\tau\big)= \exp[-\tau I(w)]\] and assuming that
$I(w)\geq 0$ is strictly convex with minimum at $\overline{w},
I(\overline{w})=0$ and which, from the EFS
\eqref{eq:theorem_workEFT}, satisfies $I(-w) - I(w) =\beta w$.
Let  $w^\star$ be the solution of $I'(w)=\beta$. In case the rate
function $I(w)$ is symmetric around $w=\overline{w}$, then
$w^\star=3\overline{w}$.\\

Under these assumptions, we will argue in Section
\ref{ssec:proof-HeatFT} that the following holds true in
general:\\

Consider, as in \eqref{FS-heat}, for $q\geqslant 0$,
\begin{equation}
\label{eq:heat-Jq} f(q) \equiv \lim_{\tau \rightarrow +\infty}
\frac 1{\tau} \log \frac{\Prob(Q_\gamma = \tau q)}{\Prob(Q_\gamma
= - \tau q)}
\end{equation}
Then,
\begin{displaymath}\label{heatfs}
f(q) = \left\{ \begin{aligned}
&\beta q && \qquad \textrm{for }0\leqslant q \leqslant \overline{w} \\
&\beta q - I(q) && \qquad \textrm{for } \overline{w}\leqslant q\leqslant w^\star \\
&\beta w^{\star} - I(w^\star) && \qquad \textrm{for }q \geqslant w^\star
\end{aligned} \right.
\end{displaymath}
%\end{enumerate}
The antisymmetry of $f(q) = - f(-q)$ fixes the behavior for $q < 0$.\\

\begin{figure}[h]
\includegraphics[scale=0.75]{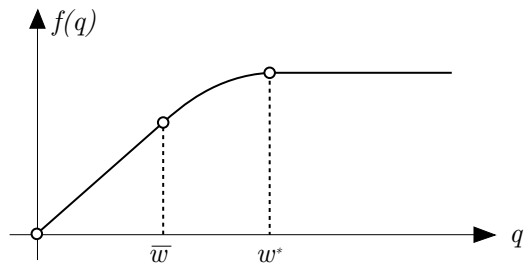}
\caption{Deviations from the fluctuation symmetry \eqref{FS-heat} for
the heat per unit time $q$. The function $f(q)$ is defined in
\eqref{eq:heat-Jq}. For small values of $q$, $f(q)$ is linear so
the standard fluctuation theorem is recovered. Between
$\overline{w}$ and $w^\star$, the behavior is determined
by the large deviation rate-function $I(q)$ of the work. The
function $f(q)$ saturates for large $q$.}
\end{figure}

 As an example, take
$I(w) = \beta(w-\overline{w})^2/4\overline{w}$ as is the case for
\eqref{eq:potential-CohenVanZon}, see \cite{CohenVanZon}. Then
 $w^\star=3\overline{w}$ and we
have three regimes.  A first linear regime where we see the usual
symmetry \eqref{eq:theorem_workEFT} for $0<q\leqslant \overline{w}$,
then a quadratic regime for $\overline{w} < q\leqslant 3\overline{w}$ which
saturates to a fixed value for $q\geqslant 3\overline{w}$.  Under our
assumptions, we have now a general expression for the
symmetry of the heat fluctuations, extending the results in
\cite{CohenVanZon} quite beyond the harmonic case
\eqref{eq:potential-CohenVanZon}.\\

A more probabilistic interpretation and a toy-calculation is
presented in Appendix II.

% ===================================================================
% DISCUSSION
% ===================================================================

%\newpage
\section{Experiments and numerical work}\label{disc}

\subsection{Simulations}
\label{ssec:numerics}

In the previous sections, we have given conditions on the protocol
and on the potential for the work to follow an EFS
\eqref{eq:theorem_workEFT}. We now argue via numerical examples
that our sufficient conditions are also close to being necessary.
To that aim, we have simulated the Langevin motion of the particle
by means of an Euler-Maruyama scheme. The time interval $[0,\tau]$
is divided into $n$ parts $dt = \tau/n$, and the evolution of the
system takes place via discrete states $x_i$ ($i=0,1,2,\ldots,n$)
connected by finite $dt$ steps,
\[
x_{i+1} = x_i - {\partial U(x_i,\gamma_i) \over \partial x} dt +
\sqrt{2 dt} B_i
\]
where $B_i$ is a random number drawn from a normal distribution, and we have
set $\beta=1$. The work \eqref{eq:definition-work} is calculated through
\begin{displaymath}
W = - \sum_{i=0}^{n-1} U'(x_{i} - \gamma_{i}) ( \gamma_{i+1} - \gamma_{i})
\end{displaymath}

\begin{figure}[t]
\begin{center}
\includegraphics[angle=0,width=8.0cm]{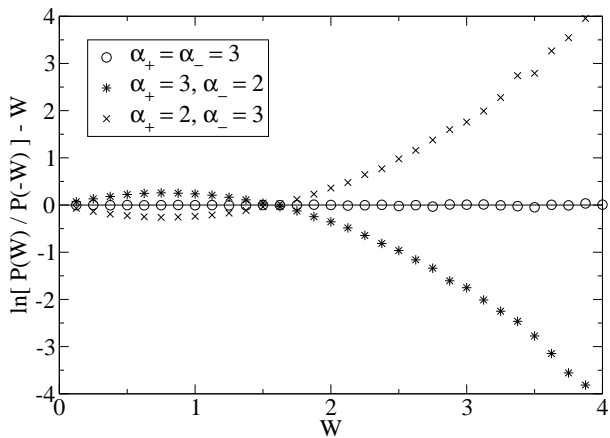}
\end{center}
\caption{Plot of the deviations from the EFS
\eqref{eq:theorem_workEFT}, for several values of the exponents
$\alpha_+$ and $\alpha_-$ in (\ref{eq:UA}). A potential which is
dragged with constant velocity $v=1$ is considered: the EFS is
verified for the symmetric potential (here we chose $\alpha_+ =
\alpha_- = 3$), while it is not observed for asymmetric
potentials. Parameters are $\tau=1$ and $dt=10^{-3}$.
\label{fig:asym-pot} }
\end{figure}

We consider the asymmetric potential, for $\alpha_+,\alpha_- > 0$,
\begin{equation}
U_t(x) = U(x - \gamma_t) = \left\{
\begin{aligned}
\frac{|x - \gamma_t|^{\alpha_+}}{\alpha_+} && \textrm{for $x \geqslant \gamma_t$}\\
\frac{|x - \gamma_t|^{\alpha_-}}{\alpha_-} && \textrm{for $x < \gamma_t$}
\end{aligned}
\right.
\label{eq:UA}
\end{equation}
The first case examined is where the potential above is moved with
 a linear protocol $\gamma_t = t$ for $t>0$.
At $t=0$, we generate equilibrated configurations, sampled with a
usual Markov chain and a Metropolis criterion. First we chose a
generic (non-harmonic) symmetric potential, with $\alpha_+ =
\alpha_- = 3$, for which we expect the EFS
\eqref{eq:theorem_workEFT} to hold. That is confirmed in figure
\ref{fig:asym-pot}, in which we plot the difference
\begin{displaymath}
\ln \left[\frac{ P(W = w) }{P(W = -w)} \right] - w
\end{displaymath}
between the lefthand-side and the righthand-side in
\eqref{eq:theorem_workEFT}. Indeed, there are no noticeable
deviations from zero for the case of a symmetric potential. On the
other hand, in the same figure, the results found for asymmetric
potentials are not in agreement with the EFS.  In that case the
conditions of Proposition \ref{prop:cases} are not verified. Note
that the symmetric deviations from the origin found for the
choices ($\alpha_+ = 3, \alpha_-=2$) and ($\alpha_+ = 2,
\alpha_-=3$) represent an indirect verification of the Crooks
relation,
see further in Section \ref{proof-CFT}.\\

%%%%%%%%%%%%%%%%%%%%%%%%%%%%%%%%%%%%%%%%%%%%%%%%%%%%%%%%%%%%%%%%%%%
%%%%%%%%%%%%%%%%%%%%%%%%%%%%%%%%%%%%%%%%%%%%%%%%%%%%%%%%%%%%%%%%%%%

\begin{figure}[t]
\begin{center}
\includegraphics[angle=0,width=8.0cm]{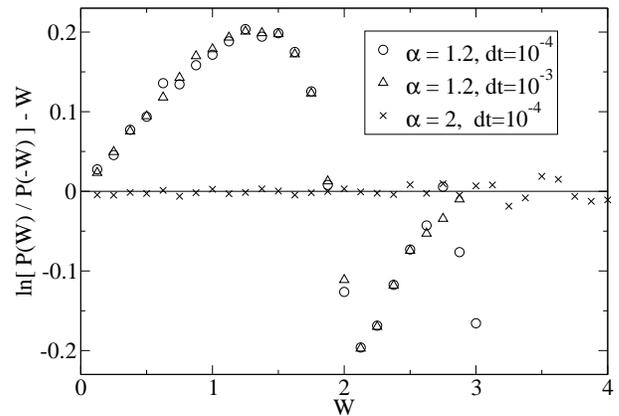}
\end{center}
\caption{Plot of the deviations from the EFS
\eqref{eq:theorem_workEFT}, for symmetric potentials [$\alpha_+ =
\alpha_-\equiv\alpha$ in (\ref{eq:UA})] and spatially translated
with protocol $\gamma_t =t+t^4$. We chose $\tau=1$ and
$dt=10^{-4}$. The fluctuation theorem is verified for the harmonic
potential, while it is not valid for a symmetric potential with
exponent $\alpha_+ = \alpha_- = 1.2$. For the latter potential, a
simulation with $dt=10^{-3}$ shows that numerical approximation is
negligible. \label{fig:asym-proto} }
\end{figure}

\begin{figure}[t]
\begin{center}
\includegraphics[angle=0,width=8.0cm]{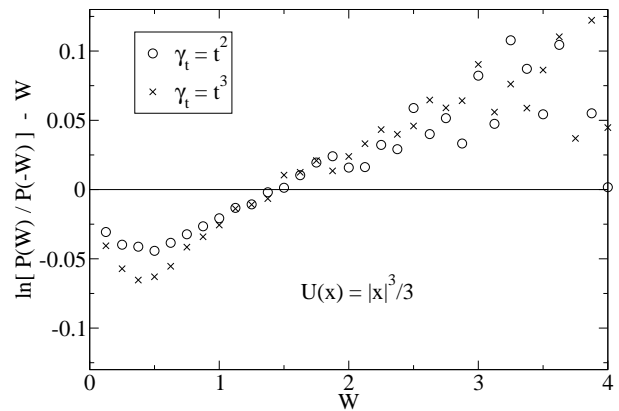}
\end{center}
\caption{Plot of the deviations from the EFS
\eqref{eq:theorem_workEFT}, for the symmetric potential $U(x) = |x|^3 / 3$ [$\alpha_+ =
\alpha_-=3$ in \eqref{eq:UA}]
and spatially translated with protocol $\gamma_t =t^2$ and $\gamma_t=t^3$. We chose $\tau=1$ and
$dt=10^{-4}$. Since both protocols are neither symmetric or antisymmetric, the EFS is indeed not expected
to hold.}  \label{fig:asym-proto2}
\end{figure}

In figure \ref{fig:asym-proto}, one sees again how our conditions
in Proposition \ref{prop:cases} are necessary.  This time we take
a protocol that lacks the suitable temporal symmetries, like
$\gamma_t = t+t^4$. The EFS is then not verified even for a
symmetric potential as in (\ref{eq:UA}) with $\alpha_+ = \alpha_-
(\neq 2)$. However, as expected, the simulation of the special
case of the harmonic potential $\alpha_+ = \alpha_- = 2$ obeys the
conclusion of Proposition \ref{prop:harmonic}. Similar conclusions
are drawn from figure \ref{fig:asym-proto2}.

%%%%%%%%%%%%%%%%%%%%%%%%%%%%%%%%%%%%%%%%%%%%%%%%%%%%%%%%%%%%%%%%%%%
%%%%%%%%%%%%%%%%%%%%%%%%%%%%%%%%%%%%%%%%%%%%%%%%%%%%%%%%%%%%%%%%%%%

%%%%%%%%%%%%%%%%%%%%%%%%%%%%%%%%%%%%%%%%%%%%%%%%%%%%%%%%%%%%%%%%%%%
%%%%%%%%%%%%%%%%%%%%%%%%%%%%%%%%%%%%%%%%%%%%%%%%%%%%%%%%%%%%%%%%%%%

\subsection{Experiments}

Previous experiments to test the fluctuation theorem for
nonequilibrium systems included a particle dragged in water. In
 \cite{wang},  Wang {\em et al.} consider a particle equilibrated in an optical trap
and then dragged by the trap at constant speed relative to the
surrounding water. The particle is micron-sized, the force is of
order of pico-Newton and about 500 particle trajectories were
recorded for times up to 2 seconds after initiation. The protocol
specifies the time-dependent position of the trap, approximated as
the position of the minimum in a harmonic potential with spring
constant $\kappa$.  The external force exerted on the particle is
thus $F_t(q) = -\kappa(q - \gamma_t)$.  The motion of $\gamma_t =
vt$ is about rectilinear. In a second experiment, \cite{evans},
the shape of the confining potential was changed. However, both
are examples of harmonic potentials, which we have shown to be
a very special class.\\

More recently, more general situations have been investigated, see
\cite{stuttgart,stuttgart2}. E.g. a two level system was realized
experimentally with a single defect in a diamond. When the system
is externally driven by a laser, the dissipation $R = \beta
W^{dis}$ displays non-Gaussian fluctuations. It was noticed that
integrated versions of the fluctuation theorem in their experiment
are observed only for particular protocols, in line with our
general results about the symmetric protocols (Proposition
\ref{prop:cases}). In the more recent paper \cite{stuttgart}, the
distribution of the work performed on a particle was computed for
a non-harmonic potential. Again, the time-symmetric protocol
 has been found to yield results consistent
with the EFS. Note however that our results show that a symmetric
protocol is not necessary; also the application of an
antisymmetric (e.g. linear) protocol combined with a symmetric
potential provides a verification
of the EFS (see Proposition \ref{prop:cases} and figure \ref{fig:asym-pot}).

% ===================================================================
% PROOF
% ===================================================================

%\newpage
\section{Proofs}
\label{ssec:proofs}

\subsection{Exact identities (Crooks and Jarzynski relations)}
\label{proof-CFT}

The proofs of the results listed in section \ref{ssec:results} are
discussed here. The basic ingredient for approaching the
fluctuations of dissipation functions via the method of
time-reversal was already mentioned in the introduction. In
particular, for stochastic dynamics and especially those that we
study here under equation \eqref{lang}, the following relation is
known as the Crooks fluctuation theorem, see \cite{Crooks};
remember the notation around \eqref{eq:def-timereversal}:
\begin{lemma}
\begin{equation}
\label{eq:theorem_CFT}
\frac{\Prob_{\rho_0}(\omega|\gamma)}{\Prob_{\rho_{\tau}}( \Theta
\omega | \Theta \gamma)} = \exp\big(\beta \left(
W_{\gamma}(\omega) - \Delta F\right)\big)
\end{equation}
\end{lemma}

\begin{proof}
Using the Girsanov formula \cite{Girsanov}, the probability density
$\Prob_{\rho_0}(\omega|\gamma)$ on trajectories can be expressed
in terms of the potential. Remember that the reference measure is
associated to the $U = 0$ case (pure Brownian trajectories),
starting from $\rho_0$,
\begin{align}\label{girs}
&\Prob_{\rho_0}(\omega|\gamma) \nonumber\\
 & = \exp\Bigg[ -\frac{\beta}{2} \int_0^{\tau} dx_t \circ \frac{\partial U_t}{\partial x}(x_t) + ST \Bigg]  \nonumber
    \\
 & = \exp\Bigg[ \frac{\beta Q_{\gamma}(\omega)}{2}  + ST
  \Bigg]
\end{align}
where
\begin{displaymath}
ST = \frac{\beta}{4} \int_0^{\tau} dt \bigg[ \frac{\partial^2 U_t}{\partial x^2}(x_t) - \bigg( \frac{\partial U_t}{\partial x}(x_t) \bigg)^2 \bigg]
\end{displaymath}
The ratio of time-forward and time-backward probabilities can then
be computed by using
\begin{equation}\label{using}
 Q_{\Theta \gamma}(\Theta \omega) = - Q_{\gamma}(\omega)
\end{equation}
\begin{displaymath}
\Theta (ST) = ST
\end{displaymath}
That leads to
\begin{align*}
\frac{\Prob_{\rho_0}(\omega|\gamma)}{\Prob_{\rho_{\tau}}(\Theta
\omega|\Theta \gamma)} & = \frac{\rho_0(x_0)}{\rho_{\tau}(x_{\tau})}
\exp\big[ \beta Q_{\gamma}(\omega) \big] \\
& = \exp\big[ \beta \Delta
U - \beta \Delta F + \beta Q_{\gamma}(\omega) \big]
\end{align*}
which is \eqref{eq:theorem_CFT} since $\Delta U = -Q_\gamma +
W_\gamma$.
\end{proof}

From the Crooks relation \eqref{eq:theorem_CFT} follows easily
the so called Jarzynski relation \cite{Jarzynski_PRL97}. In our context, this is the
normalisation of the probability distribution,
\begin{displaymath}
1 = \bigg\langle \frac{\Prob_{\rho_{\tau}}( \Theta \omega | \Theta \gamma)}{\Prob_{\rho_0}(\omega|\gamma)} \bigg\rangle_{\rho_0}
 = e^{\beta \Delta F} \braket{e^{- \beta W_{\gamma}(\omega)}}_{\rho_0}
\end{displaymath}
where $\langle \cdot \rangle_{\rho_0}$ is the expectation starting
from $\rho_0$. Conclusion:
\begin{equation}\label{jarz}
e^{-\beta \Delta F} = \braket{e^{- \beta
W_{\gamma}(\omega)}}_{\rho_0}
\end{equation}
A more microscopic and physically inspired derivation of the
Jarzynski relation follows in Appendix I.

\subsection{The harmonic potential with a general protocol}
\label{sssec:harmonicpotentialgenprotocol}
For the harmonic potential \eqref{eq:potential-quadratic},
all protocols $\gamma$ lead to an exact fluctuation symmetry for the work. The proof
can easily be generalized to other quadratic forms of the potential where for example the
protocol works multiplicatively (e.g. $U_t(x) = U(\gamma_t x)$).\\

From the definition of work \eqref{eq:definition-work}, it is easy
to see that the distribution of the work is Gaussian in the case
of a harmonic potential.\\

\begin{proof} [Proposition \ref{prop:harmonic}]
The free energy difference \eqref{eq:deltaF} satisfies $\Delta F =
0$ for all possible protocols $\gamma_t$. If the distribution of
the work is Gaussian with mean $\overline{w}$ and variance
$\sigma^2$, the expectation value in \eqref{jarz} can be computed
explicitly:
\begin{align*}
1 = \braket{e^{- \beta W_{\gamma}(\omega)}} & =
\exp\Bigg[{\frac{1}{2\sigma^2}(-2\overline{w} \sigma^2 \beta + \sigma^4
\beta^2)}\Bigg]
\end{align*}
Thus, necessarily, $\overline{w} = \frac{1}{2} \sigma^2 \beta$.\\

Finally, it is easy to check that a Gaussian random variable whose
mean $\overline{w}$ and variance $\sigma^2$ are related by
$\overline{w} = \frac{1}{2} \sigma^2 \beta$ satisfies
\eqref{har}.\end{proof}

\subsection{Work EFS}
\label{ssec:proof-EFTwork}

 By applying property
\eqref{eq:definition_s} of the involution $s$ to the numerator and
denominator of the Crooks relation \eqref{eq:theorem_CFT}, we
find:
\begin{align*}
\exp\big[ \beta \big( W_{\gamma}(\omega) - \Delta F \big) \big] &
=
\frac{\Prob_{\rho_0}(\omega|\gamma)}{\Prob_{\rho_{\tau}}(\Theta
\omega|\Theta \gamma)} \\
 & = \frac{\Prob_{\rho_\tau}(s \omega|\Theta \gamma)}
{\Prob_{\rho_0}(\Theta s \omega| \gamma)}\\
& = \exp\big[ - \beta \big( W_{\gamma}(s \Theta \omega) + \Delta
F \big) \big]
\end{align*}
Hence,
\begin{equation}\label{timerevwork}
W_{\gamma}(s \Theta \omega) = - W_{\gamma}(\omega) +  2 \Delta F
\end{equation}
From the first law combined with \eqref{using} one also concludes
that $W_{\Theta \gamma}(\Theta \omega) = -W_{\gamma}(\omega)$.
\begin{proof} [Theorem \ref{theorem:efs-work}]
Let us explicitly denote the dependence of the dynamics on the
protocol $\gamma$ by writing $\Prob_{\rho_0}(W_\gamma=w|\gamma)$
for the density of the work $W_\gamma$. By \eqref{eq:theorem_CFT}
\begin{align*}
\Prob_{\rho_0} & \big( W_{\gamma}(\omega) = w' | \gamma \big)\\ & =
e^{\beta(w' - \Delta F)}\, \Prob_{\rho_\tau} \big(
W_{\gamma}(\Theta\omega) = w' | \Theta\gamma \big)
\end{align*}
By \eqref{eq:definition_s}:
\begin{align*}
\Prob_{\rho_\tau} & \big( W_{\gamma}(\Theta\omega) = w' | \Theta\gamma\big) \\
 &=\Prob_{\rho_0}(W_\gamma(\Theta s \omega)=w'|
\gamma)
\end{align*}
As a consequence, via \eqref{timerevwork}
\begin{align*}
  \Prob_{\rho_0} & \big(
W_{\gamma}(\omega) = w' | \gamma \big) \\& = e^{\beta(w' - \Delta
F)}\, \Prob_{\rho_0}(W_\gamma(\omega)=2 \Delta F - w'| \gamma)
\end{align*}
Substituting $w'=w + \Delta F$, we find the EFS
\eqref{eq:theorem_workEFT} as required. \end{proof}

\begin{proof} [Proposition \ref{prop:cases}]
Suppose first a symmetric protocol $\Theta \gamma = \gamma$ and
hence $\gamma_{\tau} = \gamma_0$,
\begin{displaymath}
U_0(x) = U_\tau(x) \qquad \Rightarrow \qquad \rho_0(x) =
\rho_{\tau}(x)
\end{displaymath}
with $\rho_t$ the distribution \eqref{eq:definition rho}. Choosing
the identity operator as the involution $s = \mathbbm{1}$, i.e.,
$s\omega = \omega$, we find that \eqref{eq:definition_s} is
obviously satisfied.\\

 For antisymmetric protocols $\gamma_{\tau-t} = X-\gamma_t$
 with $X=\gamma_0+\gamma_\tau$ we restrict
ourselves to symmetric potentials of the form \eqref{gen}. Observe
then that
\[
 U(X-x-\gamma_t) = U(-x+\gamma_{\tau-t}) = U(x-\gamma_{\tau-t})
\]
which, for $t=0$,  implies $\rho_0(X-x) = \rho_\tau(x)$.  Choose
therefore the involution $s$ in \eqref{eq:definition_s} as the
flip $s \omega = X-\omega$, in the sense that $s(\omega)_t=X-x_t$
for $\omega=(x_t)$. Then, by simple inspection from \eqref{girs},
again by using that the potential $U$ is even, we get the equality
$\Prob_{\rho_0}(s\omega|\gamma) =
\Prob_{\rho_\tau}(\omega|\Theta\gamma)$ of densities, as for
\eqref{eq:definition_s}.
\end{proof}

\subsection{Heat FT}
\label{ssec:proof-HeatFT}

We give the arguments leading to \eqref{eq:heat-Jq}.  Here we do not
give a
full proof.\\

For very large $\tau$ it is appropriate for our purpose to
consider $Q_\gamma/\tau = W_\gamma^{dis}/\tau + \Delta(F-U)/\tau$
as the sum of two independent random variables. That asymptotic
independence can be argued for on the basis of mixing properties
of the Markov diffusion process \eqref{lang}. We thus write
formally, for arbitrary $q$,
\begin{align}\label{try}
\Prob(Q_\gamma=q\tau) & = \Prob\big(W_\gamma^{dis} + \Delta(F-U)=q\tau\big) \nonumber \\
& = \int \,dw\,e^{-\tau [I(w) + J(q-w)]}
\end{align}
where $\Prob(W_\gamma^{dis}=w\tau)= \exp[-\tau I(w)],\;
\Prob(\Delta(F-U)=u\tau)= \exp[-\tau J(u)]$ in the usual sense of
the theory of
large deviations, as $\tau \uparrow +\infty$.\\

Hence, taking the logarithm of \eqref{try} and dividing by
$\tau\uparrow+\infty$ takes us to
\begin{equation}
\label{eq:heat-inf} h(q) \equiv \lim_\tau \frac 1{\tau} \log
\Prob(Q_\gamma=q \tau) = -\inf_{w}[I(w) + J(q-w)]
\end{equation}
and we want to compute $f(q) = h(q) - h(-q)$.  As $I(w)$ is the
rate function of the large deviations of the (dissipative) work,
which we assume given and satisfying the EFS
\eqref{eq:theorem_workEFT}, the only unknown is the rate function
$J$.\\

Clearly, always in the sense of large deviations,
\[
J(u) = - \lim_{\tau\uparrow +\infty} \frac 1{\tau} \log
\Prob_{\rho_0}\bigg(\frac{U(x_0) - U(x_\tau - v\tau)}{\tau} = u \bigg)
\]
Here we assume again the independence for large $\tau$, this time
between the variables $U(x_0)$ and $U(x_\tau - v\tau)$.  Since
$U_t(x)\geq 0$, if $u
> 0$, then, by this independence,
\begin{widetext}
\begin{displaymath}
J(u)= - \lim_{\tau\uparrow +\infty} \frac 1{\tau} \log
\int_u^{+\infty}dy \,e^{-\beta y\tau}\,\Prob_{\rho_0}( U(x_\tau
- v\tau)= (y-u)\tau) \simeq \beta u
\end{displaymath}
\end{widetext}
On the other hand, if $u < 0$, we have
\[
J(u) = -\lim_{\tau\uparrow +\infty} \frac 1{\tau} \log
\Prob_{\rho_0}( U(x_\tau - v\tau)= -u\tau)
\]
Now, the process $x_\tau - v\tau$ is stationary for large $\tau$:
from equation \eqref{lang}
\[
d(x_t - vt) = -U'(x_t - vt)\,dt - v\,dt + \sqrt{\frac{2}{\beta}}
\,\, d b_t
\]
so that we can expect that for large $\tau$, $x_\tau - v\tau$ is
distributed according to the Boltzmann statistics
$\exp[-\beta U(x_\tau-v\tau) - \beta v(x_\tau-v \tau)]$.  As $U(x) \geq
|x|^{1 + \varepsilon}$, we have that for $u<0$, $J(u)=-\beta u$.\\

Summarizing, in \eqref{try} we can take $J(u) = \beta |u|$. After
all, it gives the probability of finding a huge energy difference
$\Delta(F-U)\simeq u\tau$ between the initial and the final state.
It means that either $U_0(x_0)$ or $U_\tau(x_\tau)$ must be very
large, and the energy has, in
Boltzmann statistics, an exponential distribution.\\

Finally, to obtain the results from section \ref{heatfs} one must still use
that
\[
-I(w) + I(-w) =\beta w,\qquad  I'(w) + I'(-w) = -\beta
\]
so that e.g. $I'(-\overline{w}) = -\beta$.
 It is then easily seen that
$h(q) = -I(-\overline{w}) + \beta(\overline{w}+q)$ if $q\leqslant
-\overline{w}$, $h(q)=-I(q)$ if $-\overline{w}\leqslant q
\leqslant w^\star$ and $h(q)=-I(w^\star) - \beta(q-w^\star)$ if
$q\geqslant w^\star$. From these one computes $f(q) = h(q) -
h(-q)$.

% ===================================================================
% APPENDIX: Jarzinsky
% ===================================================================

\section*{Appendix I: The basis of a Jarzynski relation}

 Let $\Gamma$ be the
phase space on which we have a time-dependent dynamics \footnote{This appendix is discussed in a similar form in
\cite{leshouches} by one of the authors.} defined in
terms of invertible transformations $f_t$. One can think of a
protocol $\gamma$ that changes in discrete steps so that a phase
space point $x\in \Gamma$ flows in time $t$ to $\varphi_{t,\gamma}
x \in \Gamma$ with
\[
\varphi_{t,\gamma} = f_{t}\, \ldots f_2\, f_1, \quad
t=1,\ldots,\tau
\]
 For the reversed
protocol $\Theta \gamma$,
\[
\varphi_{t,\Theta \gamma} = f_{\tau-t+1} \,\ldots f_{\tau-1}\,
f_{\tau}
\]
We imagine a measure $\mu$ on the phase space $\Gamma$ that is
left invariant by $\varphi_{t,\gamma}$: $\mu(\varphi_{t,\gamma}^{-1} B)=\mu(B)$
for $B \subset \Gamma$. Furthermore, $\Gamma$ is equipped with an involution
 $\pi$ that also
leaves $\mu$ invariant.  We assume dynamical reversibility in the
sense that for all $t$,
\[
 f_t \, \pi = \pi \, f_t^{-1}
 \]
As a consequence, $\pi \, \varphi_{t,\Theta \gamma}^{-1} \, \pi =
f_{\tau}\,\ldots f_{\tau-t+1}$ or $\varphi_{\tau,\gamma}^{-1}\,
\pi\,\varphi_{t,\Theta \gamma}^{-1} \, \pi
=\varphi_{\tau-t,\gamma}^{-1}$.\\

Let us now divide the phase space in a finite partition
$\hat{\Gamma}$. It corresponds to a reduced description; each
element in the partition is thought to reflect some manifest
condition of the system.   The entropy is defined \`a la Boltzmann
as
\[
 S(M) = \ln \mu(M),\quad M\in \hat{\Gamma}
 \]
For example, in Hamiltonian systems one takes the Liouville measure as
the invariant measure $\mu$, and then we obtain the conventional
Boltzmann definition $S = \ln |M|$. We fix probability laws
$\hat{\rho}$ and $\hat{\sigma}$ on the elements of the partition
and we specify the initial probability measure on $\Gamma$ as
\[
 r_{\hat\rho}(A) \equiv \sum_M
\frac{\mu(A\cap M)}{\mu(M)}\,\hat\rho(M)
\]
This probability measures $A \subset \Gamma$ using $\hat{\rho}$ at
the level of the partitions $M$ of the reduced description and
using the invariant measure $\mu$ within each partition $M$. The
reduced trajectories of the system are sequences
$\omega=(M_0,M_1,,\ldots,M_\tau)$ where $M_i\in \hat{\Gamma}$,
indicating subsequent moments when the phase space point was in
the set $M_i, i=0,\ldots,\tau$.  The path-space measure
$P_{\hat\rho, \gamma}$ gives the probability of
trajectories when starting from $r(\hat\rho)$ and using protocol $\gamma$.\\

The quantity of interest that measures the irreversibility in the
dynamics on the level of $\hat{\Gamma}$ is
 (see also \eqref{eq:theorem_CFT} and \cite{mn}):
 \[
R = \ln
\frac{P_{\hat\rho,\gamma}(M_0,M_1,\ldots,M_\tau)}{P_{\hat\sigma\pi,\Theta
\gamma}(\pi M_\tau,\pi M_{\tau-1},\ldots,\pi M_0)} \] The point is
that for every probability $\hat\rho$ and $\hat\sigma$ on
$\hat\Gamma$, and for all $M_0,\ldots,M_\tau \in \hat\Gamma$,
\begin{equation}\label{majar}
R = S(M_\tau) - S(M_0) - \ln \hat\sigma(M_\tau) + \ln
\hat\rho(M_0)
\end{equation}
To show \eqref{majar} we only have to look closer at the
consequences of the dynamic reversibility. By using that $\mu(B) =
\mu(\varphi_{\tau,\gamma}^{-1}\pi B)$, we have of course that
\begin{displaymath}
 \mu\Bigg[\bigcap_{t=0}^\tau \varphi_{t,\Theta\gamma}^{-1} \, \pi M_{\tau-t} \Bigg]
 = \mu\Bigg[\bigcap_{t=0}^\tau
  \varphi_{\tau,\gamma}^{-1} \,
   \pi \, \varphi_{t,\Theta\gamma}^{-1} \, \pi M_{\tau-t}\Bigg]
\end{displaymath}
but moreover, by reversibility, the last expression equals
\begin{displaymath}
\mu\Bigg[\bigcap_{t=0}^\tau \varphi_{\tau,\gamma}^{-1} \, \pi
\circ \varphi_{t,\Theta\gamma}^{-1} \, \pi M_{\tau-t} \Bigg] =
\mu\Bigg[\bigcap_{t=0}^\tau \varphi_{\tau-t,\gamma}^{-1}
M_{\tau-t}\Bigg]
\end{displaymath}
which is all that is needed.\\

As an immediate corollary, under the expectation
$P_{\hat\rho,\gamma}$
\begin{equation}\label{microjar}
\big\langle e^{-S(M_\tau) + S(M_0) + \ln \hat\sigma(M_\tau) - \ln
\hat\rho(M_0)} \big\rangle = 1
\end{equation}

 A simple choice for the system and partition takes
an isolated system where the reduced variables $M_i$ refer to the
energy of the system.   We have still the freedom to choose
$\hat\rho$ and $\hat\sigma$. Let us take $\hat\rho(M_0)=1$ where
indeed $M_0$ refers to the initial energy $E$. As final condition
we let the system be randomly distributed on the energy shell
$E'$. For these choices, in `suggestive' notation,
\eqref{microjar} becomes
\[
\ln
\frac{P_{E,\gamma}(E \rightarrow E')}{P_{E',\Theta\gamma}(E'\rightarrow E)}
= S(E') - S(E)
\]
Using that here $\Delta E = E' - E = W$ equals the work done, one
thus recovers the microcanonical analogue of the Crooks relation
\eqref{eq:theorem_CFT}, see
also \cite{VandenBroeck}. \\

The mathematically trivial identity \eqref{microjar} is the mother
of all Jarzynski relations. The way in which it gets realized as
for example an irreversible work-free energy relation depends on
the specific context or example. We can also split the system from
the environment. The reduced variables $(M_i)$ can for example be
chosen to consist of the microscopic trajectory for the system and
of the sequence of energies of the environment. For a thermal
environment at all times in equilibrium at inverse temperature
$\beta$, we thus get $S(M_\tau) - S(M_0) = \beta Q$ where $Q$ is
the heat that flowed into the reservoir.  On the other hand we can
take $\hat\rho$ and $\hat\sigma$ as equilibrium distributions, say
of the weak coupling form
\[
\hat\rho(M) = \frac{e^{-\beta U(x,\gamma_0)}}{Z_0}\,h(E)
\]
where $M=(x,E)$ combines the micro-state $x$ of the system and the
energy $E$ of the environment, $h(E)$ describes the
reservoir-distribution, $U(x,\gamma)$ is the energy of the system
with parameter $\gamma$. Similarly,
\[
\hat\sigma(M) = \frac{e^{-\beta U(x,\gamma_\tau)}}{Z_\tau}\,h(E)
\]
If we have that $h(E_0)\simeq h(E_\tau)$, i.e., that the energy
exchanges to the environment remain small compared to the
dispersion of the energy distribution in the reservoir, we get
from \eqref{microjar} in that context that
\[
\langle e^{-\beta Q -\beta U(x_\tau,\gamma_\tau) + \beta
U(x_0,\gamma_0)}\rangle_{\hat\rho} = \frac{Z_\tau}{Z_0}
\]
which is a version of the Jarzynski relation \eqref{jarz}.

% ===================================================================
% APPENDIX: Boundary terms
% ===================================================================

\section*{Appendix II: Large deviations}

 For the fluctuation symmetry we are interested in the
large deviations of $Q_\gamma/\tau$ from its average as
$\tau\uparrow+\infty$.  Such deviations can arise from two
sources.  First there are the large deviations of the work
$W_\gamma$, which however, we know,  satisfies an EFS.  Secondly
there is the possibility that $\Delta U$ also fluctuates to order
$\tau$. This second effect is responsible for the deviations from
the standard fluctuation relation \eqref{FS-heat}.  After all, an energy is typically
exponentially distributed and we can thus expect a competition with the fluctuations
of the work.\\

In order to clearly see the influence of the unboundedness of the
temporal boundary, we consider here the simplest set-up in which
deviations from the standard
fluctuation symmetry can be calculated exactly.\\

We consider a particle moving under the influence of a quadratic
potential and a random force. For each time step
$i=1,2,\ldots,\tau$ we take the work done on the particle $y_i$ to
be a random variable distributed according to a Gaussian of
average $m_i$ and variance $v_i$ \footnote{In what follows, we
will consider for convenience that $v_1=v_\tau=1$ although a more
general scenario is of course possible.}. Let us also consider the analogue
of the work \eqref{eq:definition-work} as the sum $W_\tau \equiv (y_1 +\ldots + y_\tau)$. By
construction, the work per unit time $w_{\tau} \equiv W_{\tau}/\tau$ is again
Gaussian with average $\overline{w}_{\tau}
=(m_1 + \ldots + m_\tau)/\tau$ and variance $\sigma^2_\tau = (v_1
+ \ldots + v_\tau)/\tau^2$. If $2\overline{w}_{\tau}=\sigma^2_{\tau}$, then,
automatically, the probability density function $\Prob(W_\tau
= w \tau) = \Prob(w_\tau = w)$ satisfies, for all $\tau$,
\begin{displaymath}
\frac{{\rm Prob}(w_{\tau} = w)}{{\rm Prob}(w_{\tau} = -w)} = e^{w}
\end{displaymath}
That is the (Gaussian) analogue of the exact fluctuation symmetry
\eqref{eq:theorem_workEFT} for the work
 (that we here, by the previous construction, assume from the start).\\

We now consider a new random variable (the analogue of the heat):
\begin{displaymath}
Q_\tau(w_\tau, y_1, y_\tau) \equiv W_\tau +
\eta[(y_\tau - a)^2 - (y_1-b)^2]
\end{displaymath}
where $a,b,\eta$ are real parameters, with density ${\rm
Prob}(Q_{\tau}=q\tau)$. The aim of our toy model is to compute
\begin{displaymath}
f(q)=\lim_{\tau\to\infty}\frac{1}{\tau}\ln\frac{{\rm
Prob}(Q_\tau=q\tau)}{{\rm Prob}(Q_\tau=-q\tau)}
\end{displaymath}
That can follow from $f(q)=h(q)-h(-q)$ with $h(q)$ the large
deviation rate function of the heat: $\Prob(Q_\tau = q\tau) \simeq
\exp(\tau h(q))$. The function $h(q)$ is the Legendre transform
of the generating function
\begin{displaymath}
E(t)=\lim_{\tau}\frac{1}{\tau}\ln E_\tau(t)
\end{displaymath}
with
\begin{equation}
E_\tau(t)=\frac{1}{(2\pi)^{\frac32}{\rm det}^{\frac12} C}\int dy\
e^{ tQ_\tau(y)}\ e^{-\frac12 (y-\bar{y})\cdot
C^{-1}(y-\bar{y})} \label{eq:En}
\end{equation}
where, collectively, $y=(w_\tau,y_1,y_\tau)$ and
$\bar{y}\equiv(\overline{w}_\tau,\bar{y}_1,\bar{y}_\tau)$ represent
their mean while $C=C_\tau$ corresponds to the covariance matrix
of $y$.
 Doing the
Gaussian integrals in (\ref{eq:En}) and taking the limit
$\tau\to\infty$ leads to
\begin{displaymath}
E(t)=\left\{
\begin{array}{cl}
\frac12 v t^2+t\overline{w} & {\rm if}\ t\in[-t_\star,t_\star]\\[3mm]\
+\infty                    & {\rm otherwise}
\end{array}\right.
\end{displaymath}
where $t_\star=1/2\eta, v = \lim_\tau \sigma^2_\tau = 2\overline{w} = 2\lim_\tau \overline{w}_\tau$.\\

We are now interested in evaluating the Legendre transform of the
above, $h(q)=-{\rm sup}_t[qt-E(t)]$. The location of the supremum
depends on whether $(q-\overline{w})/v$ lies within or outside the
interval $[-t_\star,t_\star]$. As a result, $h(q)$ becomes a
quadratic function within the interval
$[-vt_\star+\overline{w},vt_\star+\overline{w}]$ and a linear one outside.
For the final result for $f(q)$ one distinguishes between the
following two cases depending on the value of $\overline{w}$.
\begin{widetext}
For
$vt_\star<\overline{w}$
\begin{equation}
f(q)=\left\{\begin{array}{lll}
2qt_\star & & {\rm for} \ \ q\in[0,\overline{w}-vt_\star]\\[3mm]
-\frac{1}{2v}(q-\overline{w})^2+qt_\star-\frac12vt_\star^2+\overline{w}t_\star & &
{\rm for}\ \ q\in[\overline{w}-vt_\star,\overline{w}+vt_\star]\\[3mm]
2\overline{w}t_\star & &{\rm for}\ \ q\in[\overline{w}+vt_\star,\infty)
\end{array}
\right.
\end{equation}
while for $\overline{w}<vt_\star$ one has
\begin{equation}
\label{eq:app-small}
f(q)=\left\{\begin{array}{lll}
q & & {\rm for} \ \ q\in[0,-\overline{w}+vt_\star]\\[3mm]
-\frac{1}{2v}(q-\overline{w})^2+qt_\star-\frac12vt_\star^2+\overline{w}t_\star & &
{\rm for}\ \ q\in[-\overline{w}+vt_\star,\overline{w}+vt_\star]\\[3mm]
2\overline{w}t_\star & &{\rm for}\ \ q\in[\overline{w}+vt_\star,\infty)
\end{array}
\right.
\end{equation}
\end{widetext}
The results of section \ref{heatfs} and of \cite{CohenVanZon}, i.e.
the Gaussian case where $\beta=1$, are reproduced in \eqref{eq:app-small} by
choosing
 $\overline{w}=1$ and $t_{\star} = 1$ (i.e. $\eta=1/2$).

% ===================================================================
% BIBLIOGRAPHY
% ===================================================================

\end{document}